\documentclass[preprint,nofootinbib,amsmath,prd,aps,superscriptaddress,tightenlines,12pt]{revtex4}
\usepackage{graphicx}
\usepackage{epstopdf}
\usepackage{bm}
\usepackage{epsfig}
\usepackage{graphics}
\usepackage{xspace}
\usepackage{amsmath}

\def\OMIT#1{}

\begin{document}
\setlength\baselineskip{17pt}

\begin{flushright}
\vbox{
\begin{tabular}{l}
ANL-HEP-PR-12-68
\end{tabular}
}
\end{flushright}
\vspace{0.1cm}


\title{\bf Combining QCD and electroweak corrections to dilepton 
production in FEWZ}

\vspace*{1cm}

\author{Ye Li}
\email[]{yeli2012@u.northwestern.edu}
\affiliation{Department of Physics \& Astronomy, Northwestern University, Evanston, IL 60208, USA}
\affiliation{High Energy Physics Division, Argonne National Laboratory, Argonne, IL 60439, USA} 
\author{Frank Petriello}
\email[]{f-petriello@northwestern.edu}
\affiliation{Department of Physics \& Astronomy, Northwestern University, Evanston, IL 60208, USA}
\affiliation{High Energy Physics Division, Argonne National Laboratory, Argonne, IL 60439, USA} 


\vspace*{0.3cm}

\begin{abstract}
  \vspace{0.3cm}
  
  We combine the next-to-next-to-leading order (NNLO) QCD corrections to lepton-pair production through the Drell-Yan mechanism with the next-to-leading order (NLO) electroweak corrections within the framework of the FEWZ simulation code.  Control over both sources of higher-order contributions is necessary for measurements where percent-level theoretical predictions are crucial, and in phase-space regions where the NLO electroweak corrections grow large.  The inclusion of both corrections in a single simulation code eliminates the need to separately incorporate such effects as final-state radiation and electroweak Sudakov logarithms when comparing many experimental results to theory.  We recalculate the NLO electroweak corrections in the complex-mass scheme for both massless and massive final-state leptons, and modify the QCD corrections in the original FEWZ code to maintain consistency with the complex-mass scheme to the lowest order.  We present phenomenological results for LHC studies that include both NNLO QCD and NLO electroweak corrections.  In addition, we study several interesting kinematics features induced by experimental cuts in the distribution of photon radiation at the LHC.

\end{abstract}

\maketitle


\section{Introduction}

The Drell-Yan (DY) production~\cite{Drell:1970wh} of lepton pairs through the exchange of a $Z$-boson or virtual photon plays a critical role at the LHC. The relative ease of identifying two leptons leads to a very clean channel for detector calibration and performance studies~\cite{Haywood:1999qg}. The DY process can serve as a luminosity monitor \cite{Dittmar:1997md} because of its relatively large production rate, and can be used to perform precision electroweak (EW) measurements~\cite{Chatrchyan:2011ya} and constrain parton distribution functions (PDFs)~\cite{pdf}. Moreover, the study of hadronic dilepton production in the high invariant-mass region could reveal signs of new physics. 

All of these uses require theoretical predictions at the percent level of precision.  The largest correction to the leading-order predictions comes from quantum chromodynamics (QCD). QCD corrections to DY production up to the next-to-next-to-leading order (NNLO) in the strong coupling constant have been previously calculated, both for the inclusive cross section~\cite{Hamberg:1990np} and for 
differential quantities~\cite{Anastasiou:2003yy,Anastasiou:2003ds,Melnikov:2006di,Melnikov:2006kv,Catani:2009sm,Catani:2010en}.  We have previously implemented the fully differential corrections to the DY process in the form of the flexible parton-level simulation 
code FEWZ ({\bf F}ully {\bf E}xclusive ${\bf W}$ and ${\bf Z}$ Production) \cite{Melnikov:2006kv,Gavin:2010az,Gavin:2012kw}.  Using 
FEWZ, predictions for arbitrary kinematic distributions can be obtained, and for most such observables the estimated theoretical uncertainty from higher-order QCD effects is a few percent.  However, at this level of precision the full EW correction at next-to-leading (NLO) cannot be neglected. The NLO EW effects are known~\cite{Berends:1984qa,Berends:1984xv,Baur:1997wa,Baur:2001ze,CarloniCalame:2007cd,Dittmaier:2009cr}, and have been implemented in several public codes such as ZGRAD2~\cite{Baur:1997wa,Baur:2001ze} and HORACE~\cite{CarloniCalame:2007cd}. One major effect of the EW correction comes from photon radiation from the final-state leptons, which can lead to large logarithmic corrections sensitive to the lepton mass or calorimeter setting. The weak correction in the high-energy Sudakov regime~\cite{Ciafaloni:2000df,ciafaloni,Fadin:1999bq,Kuhn:1999nn} can also cause a significant deviation from the leading order contribution.

In this manuscript, we combine the NNLO QCD and NLO EW corrections to DY production of lepton pairs in a new version of the FEWZ code.  We rederive the EW one-loop correction using the complex mass scheme (CMS). The higher-order contributions from both the QCD and the EW theory are simply summed together to achieve ${\cal O}(\alpha_s^2)+{\cal O}(\alpha_{EW})$ accuracy.  In addition to providing the needed theoretical control over multiple sources of higher-order corrections, the inclusion of EW effects in FEWZ 
also eliminates the need to unfold photonic radiative effects before comparing data to theory, as is currently done in LHC studies.  For completeness, we also include the photon-induced dilepton production channel at the lowest order in perturbation theory if the selected PDF set comes with a photon distribution. 

The features of the upgraded FEWZ are summarized below.
\begin{itemize}
\item The user can either choose from two hard-coded schemes for the input parameters, the $\alpha(M_Z)$ or $G_{\mu} $ scheme, or specify each coupling manually as in the original FEWZ. However, if the user decides to manually input the coupling parameters, only the QED corrections will be included in order to protect gauge invariance.

\item Two different modes corresponding to either a zero or a non-zero lepton mass can be chosen. Massless leptons lead to collinear divergences, and therefore photon-lepton recombination must be applied when their separation drops below a certain detector resolution limit. If no recombination procedure is performed and massive leptons are selected, the predictions become sensitive to logarithms of the lepton mass.
\item Histograms of photonics variables, such as the photon $p_T$ and photon-lepton separation, have been added. They can be specified in the histogram configuration file.
\end{itemize}

In order to demonstrate the features of the updated FEWZ, we present numerous phenomenological results that can be 
compared to LHC data.  We first check the results of our calculation of the NLO EW corrections against previous results 
in the literature~\cite{Dittmaier:2009cr}, and find excellent agreement across a broad variety of observables.  We then proceed to present  results for DY production at the LHC that demonstrate the interplay between QCD and EW effects.  We also study the effects of photon radiation in the DY process at the LHC, and point out several interesting kinematic features 
that occur as a result of the imposed experimental cuts.  Our combination of fixed-order QCD at NNLO with the NLO EW corrections is complementary to other efforts which combine NLO QCD plus parton-shower effects with the EW corrections to $W$-boson production~\cite{Bernaciak:2012hj,Barze:2012tt}.

This manuscript is organized as follows. In Section~\ref{sec:conv}, we present our conventions, discuss different input parameter schemes, and introduce the complex-mass scheme for unstable particles. In Section~\ref{sec:corr}, we discuss in detail our implementation of the complex mass scheme and adapt the original QCD calculation to the complex-mass scheme. Numerical results are presented in Section~\ref{sec:result} and compared to the previous literature. Histograms of phenomenologically interesting observables at the LHC and the effect of different lepton identification procedures are studied in detail. We conclude in Section~\ref{sec:conc}.

\section{Setup and conventions \label{sec:conv}}

Only three parameters are needed as basic inputs in the electroweak theory in addition to the fermion masses. We study here 
two different input schemes, both of which use the $W$ and $Z$ masses as the first two input parameters. The last input parameter is given by either the Fermi constant in the $G_{\mu}$ scheme, or the effective electromagnetic coupling at the $Z$ mass in the $\alpha(M_Z)$ scheme. At tree level, the effective electromagnetic coupling can be derived from the Fermi constant as $\alpha_{G_{\mu}}=\sqrt{2} G_{\mu} M_W^2(1-M_W^2/M_Z^2)$.  We do not consider the fine structure constant as an input-parameter option because it receives large logarithmic corrections of the form $\alpha \ln m_f^2$, induced by light fermion masses in gauge boson self-energy insertions. The effective coupling $\alpha(M_Z)$ resums the above logarithms by running the effective electromagnetic coupling from the scale $Q=0$ to $M_Z$. The Fermi constant $G_{\mu}$ is derived from the effective theory describing the weak force in low energy processes. It is most precisely measured in muon decay and receives a radiative correction denoted as $\Delta r$, which contains $\Delta \alpha(M_Z)$. $\Delta r$ additionally contains $\Delta \rho$, which accounts for the running of the weak mixing angle and receives isospin-violating corrections induced by the heavy top-quark mass. The $G_{\mu}$ input parameter scheme has been shown to be the choice most stable against higher order EW corrections~\cite{Dittmaier:2009cr} and therefore is our default option here.  We list below the Standard Model parameters used in the updated FEWZ code:
\begin{eqnarray}
\label{SMparams}
G_{\mu} = 1.16637 \times 10^{-5}~\textrm{GeV}^{-2} &,& ~ \nonumber \\
\alpha(0) = 1/137.035999911 &,& \alpha(M_{Z}) = 1/128.91 ~, \nonumber\\
M_{Z,OS} = 91.1876~\textrm{GeV} &,& \Gamma_{Z,OS} = 2.4952~\textrm{GeV} ~, \nonumber\\
M_{W,OS} = 80.403~\textrm{GeV} &,& \Gamma_{W,OS} = 2.141~\textrm{GeV} ~, \nonumber\\
m_{e} = 5.1099891 \times 10^{-4}~\textrm{GeV} &,& m_{\mu} = 0.105658369~\textrm{GeV} ~,\nonumber\\
m_t=172.9~\textrm{GeV} &,& m_H=125~\textrm{GeV} ~.
\end{eqnarray}
%
The subscript $OS$ denotes the on-shell values of the masses and widths.  The fine structure constant $\alpha(0)$ is only used for the photon-induced process.  We note that the corrections are insensitive to the choice of Higgs boson mass $m_H$.  

We begin our discussion of the complex-mass by rewriting the on-shell expressions for the $W$ and $Z$ propagators in terms of the real and imaginary parts of their complex-plane poles:
\begin{equation}
\frac{1} {s-M_{V,OS}^2+i s \Gamma_{V,OS}/M_{V,OS} \theta(s)} = \frac{1} {s-M_V^2+i \Gamma_V M_V } \left(1 + \mathcal{O}( \frac{\Gamma_V}{M_V})\right),
\end{equation}
where $V$ stands for either the $W$-boson or $Z$-boson.  We have assumed massless decay products in writing the left-hand side of this equation. In the right-hand side we have identified
\begin{equation}
M_{W/Z}=\frac{M_{W/Z,OS}}{\sqrt{1+\Gamma_{W/Z,OS}^2/M_{W/Z,OS}^2}},
\Gamma_{W/Z}=\frac{\Gamma_{W/Z,OS}}{\sqrt{1+\Gamma_{W/Z,OS}^2/M_{W/Z,OS}^2}}.
\end{equation}
The correct description of unstable particles produced on resonance is usually accomplished through Dyson resummation of self-energy insertions. It unavoidably introduces a mixing of perturbative orders and ruins gauge invariance if done incorrectly. It is especially tricky for unstable particles running in loops, since the resonant term cannot be simply factored out before the loop integration is performed. The complex mass scheme is a prescription that consistently uses complex masses everywhere for unstable particles. The complex masses of the $W$ and $Z$ gauge bosons are defined by
\begin{eqnarray}
\mu_W^2 &=& M_W^2-i M_W \Gamma_W =  \frac{M_{W,OS}^2-i M_{W,OS} \Gamma_{W,OS}}{1+\Gamma_{W,OS}^2/M_{W,OS}^2} \nonumber\\
\mu_Z^2 &=& M_Z^2-i M_Z \Gamma_Z =  \frac{M_{Z,OS}^2-i M_{Z,OS} \Gamma_{Z,OS}}{1+\Gamma_{Z,OS}^2/M_{Z,OS}^2},
\end{eqnarray}
where $M_{W/Z,OS}$ and $\Gamma_{W/Z,OS}$ are the mass and width of each boson respectively; their numerical values are given in Eq.~(\ref{SMparams}). The weak mixing angle is given by $s_W^2=1-c_W^2=1-\mu_W^2/\mu_Z^2$, resulting in complex couplings of the fermions to the $Z$ and $W$ bosons. All relations following from gauge invariance are preserved because the masses are only modified by analytic continuation. Although the proof of unitarity order-by-order using the Cutkosky cutting rule no longer holds, the spurious terms are of higher order in the perturbation series as long as the width is a higher order object with respect to the mass. It should be noted that unstable particles should not be used as external particles in the computation of S-matrix elements~\cite{Denner:2006ic}.

There are other schemes which do not rely on complex parameters to cope with unstable particles in the propagator, such as the on-shell (OS) scheme and the pole scheme (PS). The OS scheme requires that a running width which vanishes at zero-momentum is kept after Dyson resummation.  Otherwise an artificial infrared divergence would appear from the contribution of the would-be Goldstone boson in the case with finite fermion mass~\cite{Dittmaier:2002nd}.  The OS scheme also requires the running width to be kept in the propagator during the loop integration in order to describe the resonance correctly without spoiling gauge invariance.    The PS takes advantage of the fact that the residue on the resonance peak is gauge invariant. The cross section in the pole scheme for the $Z$ boson can be schematically written as:
\begin{equation}
d \sigma = \frac{A_{resonant} (s=M_Z^2)}{s-M_Z^2+ i M_Z \Gamma_Z } 
+\frac{A_{resonant}(s)-A_{resonant}(s=M_Z^2)}{s-M_Z^2} + A_{non-resonant}(s).
\end{equation}
The PS and CMS only need the input of a fixed width and are therefore more convenient. Numerically the difference between the CMS and the PS, even at thresholds, has been shown to be below 0.1\%~\cite{Dittmaier:2009cr}. The PS requires the separation of the differential cross section into resonant and non-resonant piece after completing the loop calculation. The CMS requires the analytic continuation of internal masses appeared in loop integrals into the complex plane. We choose the CMS here because most of the analytic continuation has already been provided in the early literature. Moreover, the CMS generally yields smooth distributions in resonant regions. The only complication is an extra analytic continuation for two-point scalar integrals, since the renormalization procedure in the CMS requires complex masses to be assigned to the originally real-valued momenta. 

\section{The NLO EW correction to the DY process  \label{sec:corr}}

We discuss here the technical details of our calculation of the NLO EW corrections to lepton-pair production.  We begin by separating the gauge-invariant QED subset of the correction from the full EW result.  The photonic corrections can be further split into three distinct terms: initial-state terms, final-state terms, and initial-final interference terms. Each piece is identified through coupling combinations of quark and lepton electric charges: $Q_q^2$, $Q_l^2$ and $Q_q Q_l$ respectively. We rederive the photonic correction with finite lepton mass and zero lepton mass separately, and cross-check that the results agree under appropriate choice of electromagnetic calorimeter setting if the lepton mass is small. The weak portion of the NLO EW correction consists of gauge boson self-energy insertions, weak vertex corrections, and ZZ and WW box diagrams. All fermions except for the top quark are taken to be massless in the weak corrections.

\subsection{QED corrections}

The virtual photonic corrections to vertices as well as the fermion wave function renormalization constant contain both ultraviolet (UV) and infrared (IR) divergences. QED gauge invariance guarantees the cancellation of UV singularities between them, so that no renormalization procedure is required for the one-loop QED correction. The soft singularity is cancelled completely by the real radiation process $q\bar{q} \to l\bar{l} \gamma$.

Due to the assumption of massless quarks in the initial state, a collinear singularity arises from initial-state photon radiation.  This can be absorbed into a bare PDF in complete analogy to gluon emission in QCD. We do not introduce the photonic contribution to PDF evolution, because it is suppressed by both $\mathcal{O}(\alpha)$ and the smallness of the photon PDF itself. We use the DIS scheme in constructing the PDF counterterm. The difference compared to the $\overline{\text{MS}}$ scheme is around 10\% for the initial state radiation (ISR) contribution~\cite{Baur:1997wa}. The final state radiation (FSR) also contains a collinear singularity if the lepton is assumed to be massless. A proper procedure to combine photon and leptons when two particles travel too close together leads to a cancellation of this divergence. This always occurs experimentally for the electron, since it is hard to separate a electron from its collinear photonic radiation in the electromagnetic calorimeter. For muons, such a separation is possible and the collinear singularity is instead regulated by the finite lepton mass in the form of $\alpha \ln(m_\mu)$.  A strict isolation cut on a muon will therefore lead to a considerable photonic correction to the DY cross section. The results for massless and massive leptons are equivalent to each other up to $\mathcal{O}(m_l^2/M_Z^2)$ if the same recombination procedure is applied, as the $\ln(m_l)$ dependence cancels.

The virtual box contribution in QED contains only an IR singulary, which cancels the one arising from ISR-FSR interference terms in the real radiation process. We use dimensional regularization to regulate both UV and IR poles.  Integration-by-parts (IBP) identities are used in the loop integration to reduce all tensor integrals to a few master scalar integrals via fully automated computer algorithms \cite{Anastasiou:2004vj}. All relevant scalar integrals with complex masses are calculated by hand and checked against the existing literature~\cite{'tHooft:1978xw,Hahn:1998yk,Ellis:2007qk,Denner:2010tr}. 

The soft and collinear singularities in the real radiation diagrams are extracted using a modified two cut-off phase space slicing method~\cite{Harris:2001sx}. The first cut-off $\delta_s$ defines the soft region by $E_\gamma<\delta_s \sqrt{s}/2$, where $s$ denotes the partonic center of mass energy squared.  The second cut-off $\delta_c$ defines the collinear regions with respect to each fermion (incoming quarks and outgoing leptons) by $s_{f\gamma}<\delta_c s$. By approximating the matrix element in the soft and collinear limits and integrating over only the soft and collinear phase space regions, we can obtain the IR singularities analytically as $1/\epsilon$ poles in dimensional regularization. The remaining hard non-collinear phase space can be integrated numerically in four dimensions. Given sufficiently small $\delta_s$ and $\delta_c$ parameters (terms of $\mathcal{O}(\delta_s)$, $\mathcal{O}(\delta_c)$ and $\mathcal{O}(\delta_c/\delta_s)$ are dropped), the sum of soft, collinear and hard non-collinear contributions will be independent of the cut-offs. A simply illustration of the cut-off method is as follows:
\begin{eqnarray*}
\int^1_0 \frac{f(x)}{x^{1+\epsilon}} &=& \int^\delta_0 \frac{f(x)}{x^{1+\epsilon}} + \int^1_\delta \frac{f(x)}{x^{1+\epsilon}}\\
&=& \int^\delta_0 \frac{f(0)}{x^{1+\epsilon}} + \int^1_\delta \frac{f(x)}{x} + \mathcal{O}(\epsilon) + \mathcal{O}(\delta) \\
&=& \left[- \frac{f(0)}{\epsilon} + f(0) \ln(\delta) \right] + \int^1_\delta \frac{f(x)}{x},
\end{eqnarray*}
\noindent
where we have dropped the $\mathcal{O}(\epsilon)$ and $\mathcal{O}(\delta)$ terms as they are infinitesimal parameters. The first term corresponds to soft and collinear contributions which are regulated via dimensional regularization. The second term corresponds to the hard non-collinear contribution which is limited by the cut-off near the phase space boundary. We can easily conclude from observation of the above formula that extremely small cut-off parameters would lead to numerical instability. To avoid such issues and reduce artificial parameter dependence in FEWZ, we further work out the analytic dependence of the cut-off parameters by introducing counterterms, shown in the example below:
\begin{eqnarray*}
\int^{x_0}_0 \frac{f(x)}{x^{1+\epsilon}} &=& \left[- \frac{f(0)}{\epsilon} + f(0) \ln(\delta) \right] + \int^{x_0}_\delta \frac{f(x)}{x}\\
&=&  \left[- \frac{f(0)}{\epsilon} + f(0) \ln(\delta) \right] + \int^{x_0}_\delta \frac{f(x)}{x} - \int^{x_0}_\delta \frac{f(0)}{x}+ \int^{x_0}_\delta \frac{f(0)}{x}\\
&=&  \left[- \frac{f(0)}{\epsilon} + f(0) \ln(\delta) \right] + \int^{x_0}_0 \frac{f(x)}{x} -  \int^{x_0}_0 \left(\frac{f(0)}{x} - \frac{f(0)}{x_0} \ln(\frac{x_0}{\delta})\right).
\end{eqnarray*}
The third term explicitly depends on the cut-off parameter and is our desired counterterm. It essentially becomes a subtraction method \cite{Frixione:1995ms,Catani:1996vz}.  The approximate matrix elements in the soft and collinear regions are integrated over 2-body phase space and cancel IR singularities from virtual diagrams, which can be interpreted as converting the $1/\epsilon$ pole in dimensional regulation to the cut-off regulator.  The corresponding counterterms render the integrand of the 3-body phase space finite in the soft and collinear limits, and the hard non-collinear region can be extended to the full phase space. Agreement is found numerically before and after applying the subtraction procedure and the dependence on cut-off parameters indeed vanishes. The derivation of the couterterms can be fully automated after choosing a proper parameterization of the 3-body phase space, which is done in four space-time dimensions as the cut-off now regulates IR singularities.

\subsection{Weak Correction}

The weak correction is decomposed into self-energy insertions of $\gamma\gamma$, $ZZ$ and $\gamma Z$ mixing contributions, corrections to the $l \bar{l} \gamma/Z$ and $q \bar{q} \gamma/Z$ vertices, and $ZZ$ and $WW$ box contributions. The leading contribution comes from two pieces: the running of the electromagnetic coupling from the scale $Q=0$ to the scale of the hard interaction, and the isospin violation due to the large mass splitting between top and bottom quarks. The former receives contributions largely from light fermion loops in the photon self energy insertion $\hat{\Pi}_{f\not=t}(M_Z^2) $. The latter can be mostly accounted for by the difference between the self-energy insertions of the $W$ and $Z$ bosons, $\Delta  \rho$.  We adopt most of the results for the weak correction directly from Refs.~\cite{Dittmaier:2009cr} and~\cite{Denner:1991kt}. The renormalization of the vertices and self energy insertions in the CMS is described in detail in Section 3.3 of Ref.~\cite{Dittmaier:2009cr}. Because we implement only the $\alpha(M_Z)$ and $G_{\mu}$ input parameter schemes, the logarithmic dependence on the small fermion mass cancels out in the coupling renormalization, as demonstrated below.
\begin{itemize}

\item $\alpha(M_Z)$ scheme:
\begin{eqnarray}
\frac{\delta g_{f\bar{f}\gamma/Z}}{g_{f\bar{f}\gamma/Z}} \supset \frac{\delta e(M_Z)}{e(M_Z)} &=&\delta Z_e - \frac{1}{2} \Delta \alpha (M_Z) \nonumber \\
&=& - \frac{s_W}{c_W} \frac{\Sigma^{\gamma Z}_{T}(0)}{\mu_Z^2} + \frac{1}{2} Re \left\{ \frac{\Sigma^{\gamma\gamma}_{T,f\not=t}(M_Z^2)}{M_Z^2} \right\};
\end{eqnarray}

\item $G_{\mu}$ scheme:
\begin{eqnarray}
\frac{\delta g_{f\bar{f}\gamma/Z}}{g_{f\bar{f}\gamma/Z}}  \supset \frac{\delta e_{G_{\mu}}}{e_{G_{\mu}}} &=& \delta Z_e - \frac{1}{2} \Delta r\nonumber \\
&=& - \frac{s_W}{c_W} \frac{\Sigma^{\gamma Z}_{T}(0)}{\mu_Z^2} + \frac{c_W^2}{2 s_W^2} \left( \frac{\Sigma^{ZZ}_{T}(\mu_Z^2)}{\mu_Z^2} - \frac{\Sigma^{W}_{T}(\mu_W^2)}{\mu_W^2}  \right) - \frac{\Sigma^{W}_{T}(0)-\Sigma^{W}_{T}(\mu_W^2)}{2\mu_W^2} \nonumber \\
&& - \frac{c_W}{s_W}\frac{\Sigma^{\gamma Z}_{T}(0)}{\mu_Z^2} - \frac{\alpha}{8\pi s_W^2}\left( 6+\frac{7-4s_W^2}{2s_W^2}\ln(c_W^2)\right).
\end{eqnarray}

\end{itemize}
Both behave well in the limit of massless fermions. We have suppressed chirality indices in our notation for simplicity. Another logarithmic dependence on light fermion mass comes from the photon and $Z$ boson wave function renormalization constants. However, they appear in both the vertex and self-energy renormalizations and nicely cancel  each other. We can therefore safely neglect all light quark and lepton masses in the computation of the weak correction. Because the top quark remains the only massive particle, the CKM matrix can be treated as unity and does not need to be renormalized.

Another complication of renormalization in the CMS is due to the analytic continuation of the complex momentum appearing in two-point scalar integrals, required in the calculation of gauge boson self energy insertions. An alternative procedure has been proposed in Ref.~\cite{Denner:2005fg} based on the expansion around real-valued masses. The error from the expansion is of $\mathcal{O}(\Gamma_Z/M_Z)$ for the $Z$ boson self-energy insertion. Special care must to be taken for charged or colored particles like the $W$ boson. The expansions break down in the presence of photon or gluon exchange, which generates terms like $(s-\mu_W^2) \ln(s-\mu_W^2)$. We instead adopt the direct approach of performing the analytic continuation. We first note that except for light fermions whose masses are assumed to be zero, all particles in the $W/Z$ boson self-energy graphs have masses greater than or equal to $M_{W/Z}$. Since the internal propagators of these particles cannot go on-shell when the incoming energy is at the mass of the $W$ or $Z$ boson, there should be no imaginary part from the loop integration. This implies that no branch cut is crossed in the integration, and the analytic continuation can be performed using the analytic expression derived for real-valued momenta.  In contrast, massless fermions running in the self-energy graphs go on-shell for any time-like momentum, and the integration contour moves away from the branch cut in the direction specified by the Feynman prescription. A complex mass for the incoming momentum squared leads to the opposite pole treatment from the Feynman prescription. As we will show later, a simple addition of $2 \pi i$ to the original result will fix this issue. A detailed derivation can be found in the Appendix \ref{sect:acB0}.

\subsection{Higher-order QCD corrections in the CMS}

\begin{figure}[h]
\begin{center}
\includegraphics[scale=0.6,angle=0]{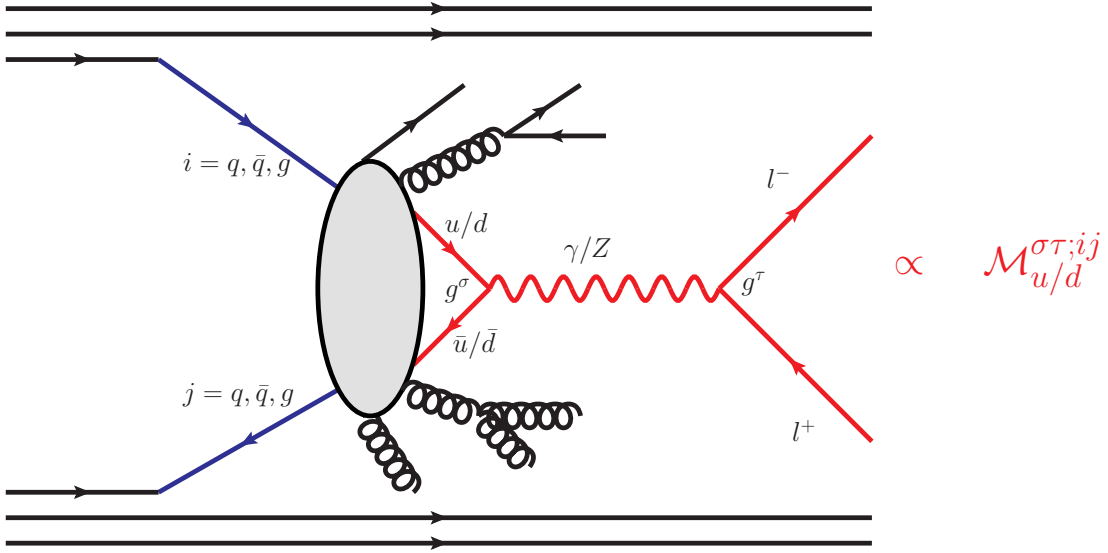}
\caption{Graphical representation of the QCD corrections to lepton pair production.}
\label{QCDcorrDY}
\end{center}
\end{figure}

The application of the CMS, and specifically the complex coupling, leads to a different LO expression than the original result implemented in FEWZ.  Although the difference is numerically small, we modify the QCD corrections in FEWZ in order to achieve theoretical consistency. In the limit of massless leptons, the QCD correction to dilepton production can be schematically written as:
\begin{eqnarray}
d\sigma^{\gamma\gamma}_{p\bar{p}\to l \bar{l}} &\sim& \frac {Q_u^2 Q_l^2 |\mathcal{M}_u|^2 + Q_d^2 Q_l^2 |\mathcal{M}_d|^2 + 2 Q_u Q_d Q_l^2 Re[\mathcal{M}_u\mathcal{M}_d^*]} {s^2}, \nonumber\\
d\sigma^{\gamma Z}_{p\bar{p}\to l \bar{l}} &\sim&  2 \displaystyle\sum\limits_{\sigma,\sigma',\tau=+,-} \left\{ \frac{Q_u Q_l}{s} Re\left[ \frac {g_{u\bar{u}Z}^{\sigma} g_{l\bar{l}Z}^{\tau}}{s-\mu_Z^2} \mathcal{M}_u^{\sigma \tau} \mathcal{M}_u^{\sigma' \tau *} \right]
+ \frac{Q_d Q_l}{s} Re\left[ \frac {g_{d\bar{d}Z}^{\sigma} g_{l\bar{l}Z}^{\tau}}{s-\mu_Z^2} \mathcal{M}_d^{\sigma \tau} \mathcal{M}_d^{\sigma' \tau *} \right]
 \right. \nonumber\\
&& + \left. \frac{Q_u Q_l}{s} Re\left[ \frac {g_{d\bar{d}Z}^{\sigma} g_{l\bar{l}Z}^{\tau} \mathcal{M}_u^{\sigma' \tau *}\mathcal{M}_d^{\sigma \tau} }{s-\mu_Z^2} \right] + \frac{Q_d Q_l}{s} Re\left[ \frac {g_{u\bar{u}Z}^{\sigma} g_{l\bar{l}Z}^{\tau} \mathcal{M}_u^{\sigma \tau}\mathcal{M}_d^{\sigma' \tau *} }{s-\mu_Z^2} \right]  \right\}, \nonumber\\
d\sigma^{ZZ}_{p\bar{p}\to l \bar{l}} &\sim& \displaystyle\sum\limits_{\sigma,\tau=+,-} \left\{ \frac{|g_{u\bar{u}Z}^{\sigma}|^2 |g_{l\bar{l}Z}^{\tau}|^2}{|{s-\mu_Z^2}|^2} |\mathcal{M}_u^{\sigma \tau}|^2 +  \frac{|g_{d\bar{d}Z}^{\sigma}|^2 |g_{l\bar{l}Z}^{\tau}|^2}{|{s-\mu_Z^2}|^2} |\mathcal{M}_d^{\sigma \tau}|^2 \right\} \nonumber \\
&& + 2 \displaystyle\sum\limits_{\sigma\not=\sigma',\tau=+,-} \frac{|g_{l\bar{l}Z}^{\tau}|^2}{|{s-\mu_Z^2}|^2} \left\{ Re[g_{u\bar{u}Z}^{\sigma}g_{u\bar{u}Z}^{\sigma' *} \mathcal{M}_u^{\sigma \tau}\mathcal{M}_u^{\sigma' \tau *}]
+ Re[g_{d\bar{d}Z}^{\sigma}g_{d\bar{d}Z}^{\sigma' *} \mathcal{M}_d^{\sigma \tau}\mathcal{M}_d^{\sigma' \tau *}]  \right\} \nonumber \\
&& +  2 \displaystyle\sum\limits_{\sigma,\sigma',\tau=+,-}  \frac{|g_{l\bar{l}Z}^{\tau}|^2}{|{s-\mu_Z^2}|^2} Re[g_{u\bar{u}Z}^{\sigma} g_{d\bar{d}Z}^{\sigma' *} \mathcal{M}_u^{\sigma \tau}\mathcal{M}_d^{\sigma' \tau *}],
\end{eqnarray}

\noindent
in which $d\sigma^{\gamma\gamma}_{p\bar{p}\to l \bar{l}}$, $d\sigma^{ZZ}_{p\bar{p}\to l \bar{l}}$ and $d\sigma^{\gamma Z}_{p\bar{p}\to l \bar{l}}$ are the contributions from photon, $Z$ and photon-$Z$ interference channels. $\mathcal{M}_u$ and $\mathcal{M}_d$ are matrix elements connected to the photon or $Z$ propagator through up-type quark and d-type quark vertices respectively, $\sigma$ and $\sigma'$ are chirality indices of the relevant quarks, and $\tau$ is the chirality index of the leptons. This structure is illustrated in Fig.~(\ref{QCDcorrDY}) where the incoming parton line, though plotted as fermion line, could represent either a quark or a gluon. 
We have abused the chirality index so that a plus(minus) index means the interacting fermion and anti-fermion in the vertex are right(left)-handed and left(right)-handed separately due to helicity conservation. This definition of the matrix element incorporates all QCD corrections to the initial state.  We note that the chirality indices of the quarks to which the vector boson couples are not necessarily related to the polarizations of the incoming partons labeled as $i$ and $j$. All chirality combinations are summed if the indices are suppressed, and the convolution with appropriate PDFs is omitted for simplicity.  The full result with all its structure intact can be restored by the following substitutions:

\begin{eqnarray}
|\mathcal{M}_{u/d}|^2 &=& \displaystyle\sum\limits_{\sigma,\tau=+,-} |\mathcal{M}^{\sigma \tau}_{u/d}|^2 
+ 2 \displaystyle\sum\limits_{\sigma\not=\sigma',\tau=+,-} Re[\mathcal{M}_{u/d}^{\sigma \tau}\mathcal{M}_{u/d}^{\sigma' \tau *}], \nonumber\\
Re[\mathcal{M}_u\mathcal{M}_d^*] &=& \displaystyle\sum\limits_{\sigma,\sigma',\tau=+,-} Re[\mathcal{M}_{u}^{\sigma \tau} \mathcal{M}_{d}^{\sigma' \tau *}],  \nonumber \\
|\mathcal{M}_{u/d}^{\sigma \tau}|^2 &\to& \displaystyle\sum\limits_{i,j=q,\bar{q},g} |\mathcal{M}^{\sigma \tau;i j}_{u/d}(x_1,x_2)|^2 f_{i}(x_1)  f_{j}(x_2), \nonumber\\
Re[\mathcal{M}_{u/d}^{\sigma \tau} \mathcal{M}_{u/d}^{\sigma' \tau *}] &\to& \displaystyle\sum\limits_{i,j=q,\bar{q},g} Re[\mathcal{M}^{\sigma \tau;i j}_{u/d}(x_1,x_2)\mathcal{M}^{\sigma' \tau;i j *}_{u/d}(x_1,x_2)] f_{i}(x_1)  f_{j}(x_2), \nonumber\\
Re[\mathcal{M}_{u}^{\sigma \tau} \mathcal{M}_{d}^{\sigma' \tau *}] &\to& \displaystyle\sum\limits_{i,j=q,\bar{q},g} Re[\mathcal{M}^{\sigma \tau;i j}_{u}(x_1,x_2)\mathcal{M}^{\sigma' \tau;i j *}_{d}(x_1,x_2)] f_{i}(x_1)  f_{j}(x_2).
\end{eqnarray}

In the original FEWZ, combinations of charge, vector and axial couplings were multiplied by the corresponding PDF and $\gamma/Z$ propagator, forming a luminosity function, which was then convoluted with the squared matrix elements $\mathcal{M}_{q_i}^{\sigma\tau}\mathcal{M}_{q_j}^{\sigma'\tau' *}$. Here $\sigma$, $\sigma'$, $\tau$ and $\tau'$ denote vector or axial couplings. Only the real part was kept for the squared matrix element since the luminosity function is real-valued (except for $\gamma Z$ interference terms due to the width in the $Z$ propagator; the imaginary part was neglected based on the same reason as explained below for the CMS). This is no longer true in the CMS. However, we can calculate the luminosity function with complex numbers, and take the real part in the end. By effectively neglecting the imaginary part of the squared matrix element, we change the cross section by $\mathcal{O}(\Gamma_{W/Z}/M_{W/Z}) \approx \mathcal{O}(\alpha)$. The cross section at the lowest order does not change because its matrix element is always real. We conclude that the QCD part of FEWZ is changed from the correct formula only by $\mathcal{O}(\alpha_s \alpha)$, which is beyond our approximation and can be safely neglected. In practice, we can apply complex couplings in the QCD module of FEWZ by declaring the coupling constants as complex numbers and modifying the following EW coupling combinations:

\begin{eqnarray}
\left( g^\sigma_{u\bar{u}Z/d\bar{d}Z/l\bar{l}Z} \right)^2 &\to& \left| g^\sigma_{u\bar{u}Z/d\bar{d}Z/l\bar{l}Z} \right|^2, \nonumber\\
g^\sigma_{u\bar{u}Z/d\bar{d}Z} g^{\sigma'}_{u\bar{u}Z/d\bar{d}Z} &\to& Re\left[ g^\sigma_{u\bar{u}Z/d\bar{d}Z} g^{\sigma' *}_{u\bar{u}Z/d\bar{d}Z} \right].
\end{eqnarray}
The electric charges are always real and therefore no operation is necessary for the $\gamma\gamma$ and $\gamma Z$ interference channels. The modification is not required if proper conjugation of complex coupling is carried out from the beginning of the derivation. Alternatively, we can implement it in terms of vector and axial coupling as follows:

\begin{eqnarray}
\left( g^{V/A}_{u\bar{u}Z/d\bar{d}Z/l\bar{l}Z} \right)^2 &\to& \left| g^{V/A}_{u\bar{u}Z/d\bar{d}Z/l\bar{l}Z} \right|^2, \nonumber\\
g^V_{u\bar{u}Z/d\bar{d}Z/l\bar{l}Z} g^A_{u\bar{u}Z/d\bar{d}Z/l\bar{l}Z} &\to& Re\left[ g^V_{u\bar{u}Z/d\bar{d}Z/} g^{A *}_{u\bar{u}Z/d\bar{d}Z/l\bar{l}Z} \right].
\end{eqnarray}

\section{Phenomenological results \label{sec:result}}

We are now ready to present numerical results using the updated FEWZ simulation code.  We split our phenomenological results into three sections.  We first provide a detailed comparison of the 
EW corrections implemented in our code with previous calculations in the literature, to demonstrate consistency between them.  We then present several 
representative distributions that illustrate the combined NNLO QCD and NLO EW corrections obtained using FEWZ.  In the last part we study distributions of the radiated photon, and point out several interesting kinematic features that should be observable in LHC data.

\subsection{Comparison with the previous literature}

We begin our presentation of phenomenological results by cross-checking our EW corrections against the results presented in Ref.~\cite{Dittmaier:2009cr}.  We take the top quark mass to be $m_t=174.6$ GeV and the Higgs boson mass $m_H=115$ GeV.  The $G_{\mu}$ input parameter scheme is chosen and the MRST2004QED PDF set is used to include the contribution of the photon PDF. In the presence of photon radiation, the following recombination procedure is applied before any acceptance cuts on leptons are implemented. Photons with rapidity $|\eta_{\gamma}| > 3$ are discarded as beam remnants. The separation between surviving photons and each lepton, $\Delta R_{l^{\pm}\gamma} = \sqrt{\Delta\eta_{l^{\pm}\gamma}^{2}+\Delta \phi_{\l^{\pm}\gamma}^{2}}$, is calculated, and the photon is recombined with the closest (anti-)lepton if $\Delta R_{l^{\pm}\gamma} <0.1$. The following acceptance cuts on leptons are then applied:
\begin{center}
  $p_{T,l^{\pm}} > 25$ GeV, $|\eta_{l^{\pm}}| < 2.5$.
\end{center} 
%


\begin{table}[h]
\begin{tabular}{ l | c | c | c | c | c | c }
\hline 
\hline

$M_{ll}$/GeV & $ > 50$ & $ > 100$ & $ > 200$ & $ > 500$ & $ > 1000$ & $ > 2000$ \\
\hline 
\hline

LO(DH)/pb & 738.733(6) & 32.7236(3) & 1.48479(1) & 0.0809420(6) & 0.00679953(3) & 0.000303744(1) \\
LO$_{0}$/pb & 738.789(9) & 32.723(4) & 1.483(1) & 0.0809449(8) & 0.0067993(6) & 0.0003038(1) \\
LO$_{\mu}$/pb & 738.769(9) & 32.728(4) & 1.483(1) & 0.0809451(8) & 0.0067993(6) & 0.0003037(1) \\

\hline

$\delta^{\gamma\gamma,LO}$(DH)/\% & 0.17 & 1.15 & 4.30 & 4.92 & 5.21 & 6.17 \\
$\delta^{\gamma\gamma,LO}$/\% & 0.17 & 1.15 & 4.30 & 4.92 & 5.21 & 6.18 \\

\hline 

$\delta^{QED,rec}$(DH)/\% & -1.81 & -4.71 & -2.92 & -3.36 & -4.24 & -5.66 \\
$\delta^{QED,rec}_{0}$/\% & -1.79 & -4.80 & -2.94 & -3.41 & -4.33 & -5.81 \\
$\delta^{QED,rec}_{\mu}$/\% & -1.77 & -4.78 & -2.93 & -3.41 & -4.33 & -5.83 \\

\hline

$\delta^{QED}_{\mu}$(DH)/\% & -3.34 & -8.85 & -5.72 & -7.05 & -9.02 & -12.08 \\
$\delta^{QED}_{\mu}$/\% & -3.38 & -9.09 & -5.85 & -7.22 & -9.28 & -12.47 \\

\hline

$\delta^{weak}$(DH)/\% & -0.71 & -1.02 & -0.14 & -2.38 & -5.87 & -11.12 \\
$\delta^{weak}$/\% & -0.70 & -1.02 & -0.14 & -2.38 & -5.87 & -11.11 \\

\hline 
\hline 
\end{tabular}
\caption{Cross sections of dilepton production at a 14 TeV LHC using MRST2004QED PDFs, for various cuts on the dilepton invariant mass.  The content of each row is explained in detail in the text. }
\label{LHCsigmas}
\end{table}


\begin{table}[h]
\begin{tabular}{ l | c | c | c | c | c | c }
\hline 
\hline

$M_{ll}$/GeV & $ > 50$ & $ > 100$ & $ > 150$ & $ > 200$ & $ > 400$ & $ > 600$ \\
\hline 
\hline

LO(DH)/pb & 142.7878(7) & 6.62280(3) & 0.824114(3) & 0.294199(1) & 0.01775063(5) & 0.001778465(5) \\
LO$_{0}$/pb & 142.783(1) & 6.6215(8) & 0.8241(3) & 0.294195(2) & 0.017750(2) & 0.0017789(6) \\
LO$_{\mu}$/pb & 142.783(1) & 6.6225(8) & 0.8246(3) & 0.294197(2) & 0.017749(2) & 0.017785 (6)\\

\hline 

$\delta^{\gamma\gamma,LO}$(DH)/\% & 0.15 & 0.72 & 1.54 & 1.44 & 0.83 & 0.57 \\
$\delta^{\gamma\gamma,LO}$/\% & 0.15 & 0.72 & 1.54 & 1.44 & 0.83 & 0.57 \\

\hline 

$\delta^{QED,rec}$(DH)/\% & -1.85 & -4.87 & -3.65 & -3.83 & -5.16 & -6.56 \\
$\delta^{QED,rec}_{0}$/\% & -1.82 & -4.96 & -3.70 & -3.89 & -5.29 & -6.76 \\
$\delta^{QED,rec}_{\mu}$/\% & -1.80 & -4.93 & -3.68 & -3.88 & -5.28 & -6.75 \\

\hline

$\delta^{QED}_{\mu}$(DH)/\% & -3.44 & -8.93 & -6.46 & -6.86 & -9.56 & -12.42 \\
$\delta^{QED}_{\mu}$/\% & -3.47 & -9.15 & -6.59 & -7.02 & -9.84 & -12.83 \\

\hline

$\delta^{weak}$(DH)/\% & -0.70 & -1.01 & -0.12 & -0.15 & -1.25 & -2.60 \\
$\delta^{weak}$/\% & -0.70 & -1.00 & -0.13 & -0.15 & -1.25 & -2.60 \\

\hline 
\hline 
\end{tabular}
\caption{
Cross sections of dilepton production at the 1.96 TeV Tevatron using MRST2004QED PDFs, for various cuts on the dilepton invariant mass.  The content of each row is explained in detail in the text.
}
\label{TEVsigmas}
\end{table} 

Tables~\ref{LHCsigmas} and~\ref{TEVsigmas} summarize the comparisons between our results and those of Ref.~\cite{Dittmaier:2009cr}, denoted by DH. The subscript 0 denotes results obtained in the massless lepton mode, while the subscript $\mu$ indicates results obtained using the muon mass. When the recombination procedure is applied, the results given by the muon mode and massless lepton mode should be identical, since the large logarithms  $\alpha \ln(m_l^2/M_Z^2)$ in the FSR and virtual photonic contributions cancel . Separate results for muons without recombination are given for comparison. The weak correction is generally small. For the LHC, it is enhanced in the high mass tail due to large EW Sudakov logarithms, reaching the same order as QCD and QED corrections. Results of photon-induced dilepton production are also listed, and are very small due to the suppression of photon PDF and the lack of resonance structure in this channel. However, its contribution can reach~5\% in the high energy range as pointed out in Ref.~\cite{Dittmaier:2009cr}. We find agreement with the numerical results of Ref.~\cite{Dittmaier:2009cr}. The small difference at the 0.1\% level for the photonic correction can be explained by the inclusion of the photon PDF in a redefined quark PDF, which was 
performed in Ref.~\cite{Dittmaier:2009cr} but not here.

\begin{figure}[!ht]
\begin{minipage}[b]{3.0in}
  \includegraphics[width=3.0in]{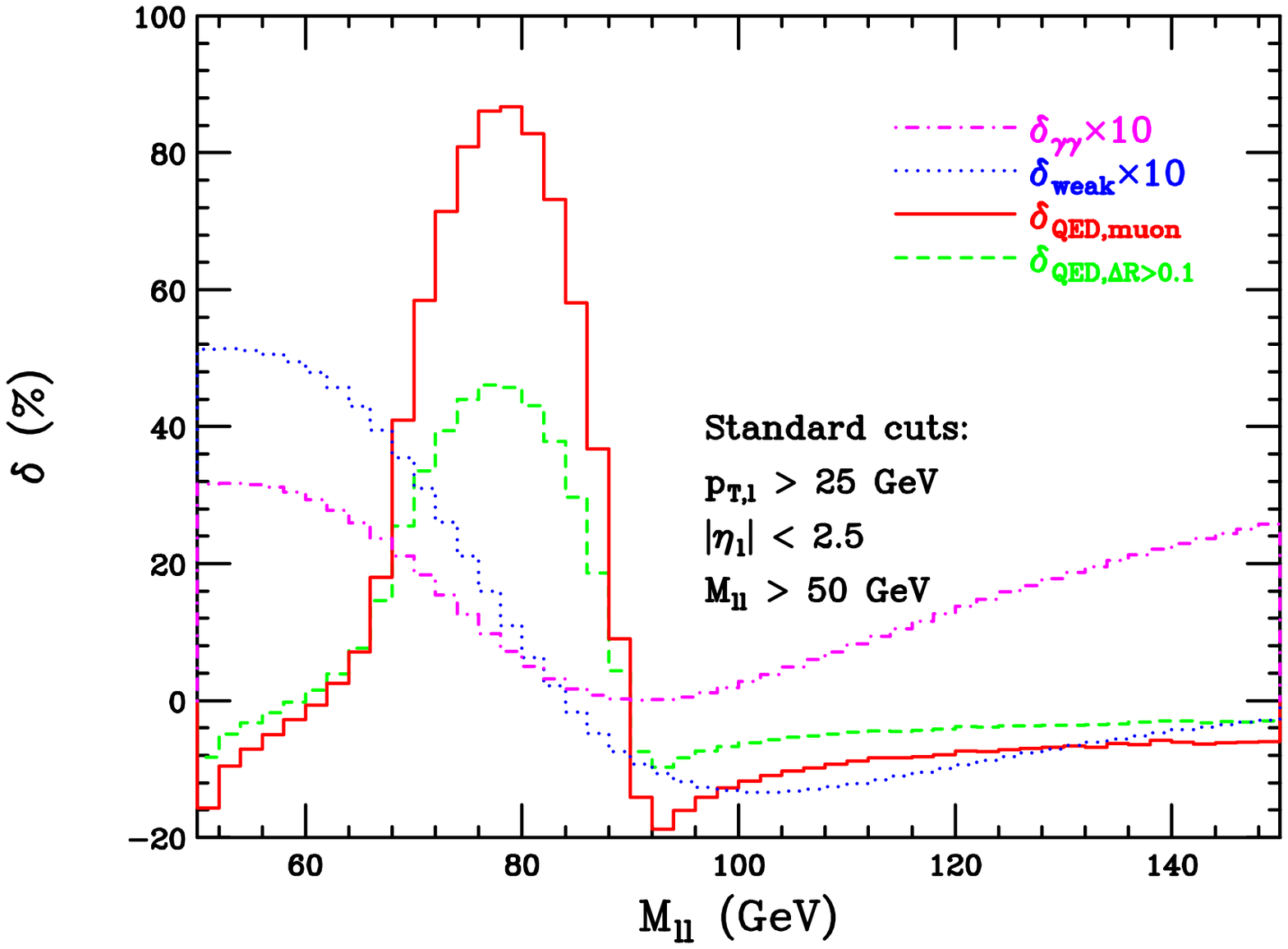}
\end{minipage}
\begin{minipage}[b]{3.0in}
  \includegraphics[width=3.0in]{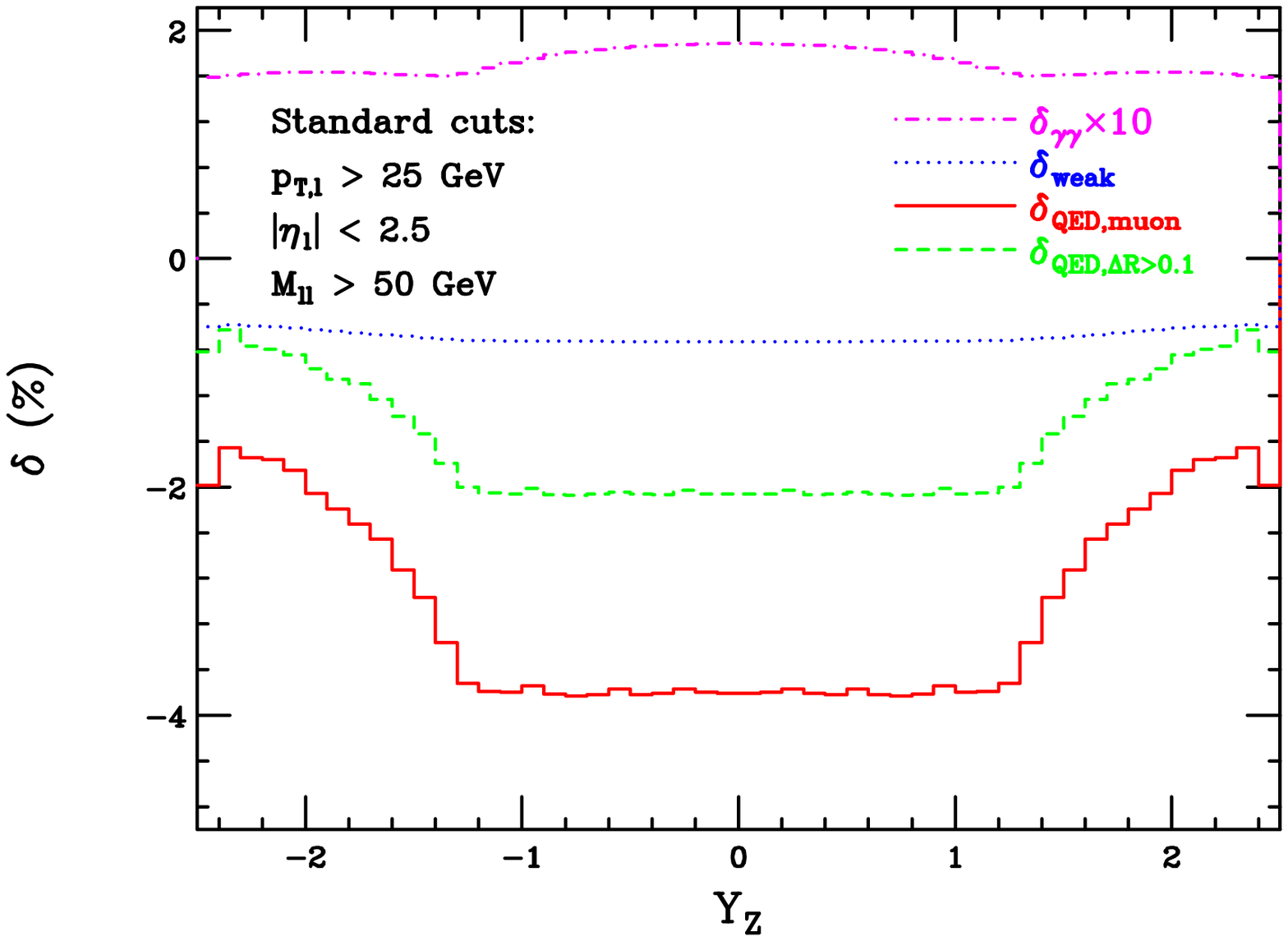}
\end{minipage}
\begin{minipage}[b]{3.0in}
\vspace{1cm}
  \includegraphics[width=3.0in]{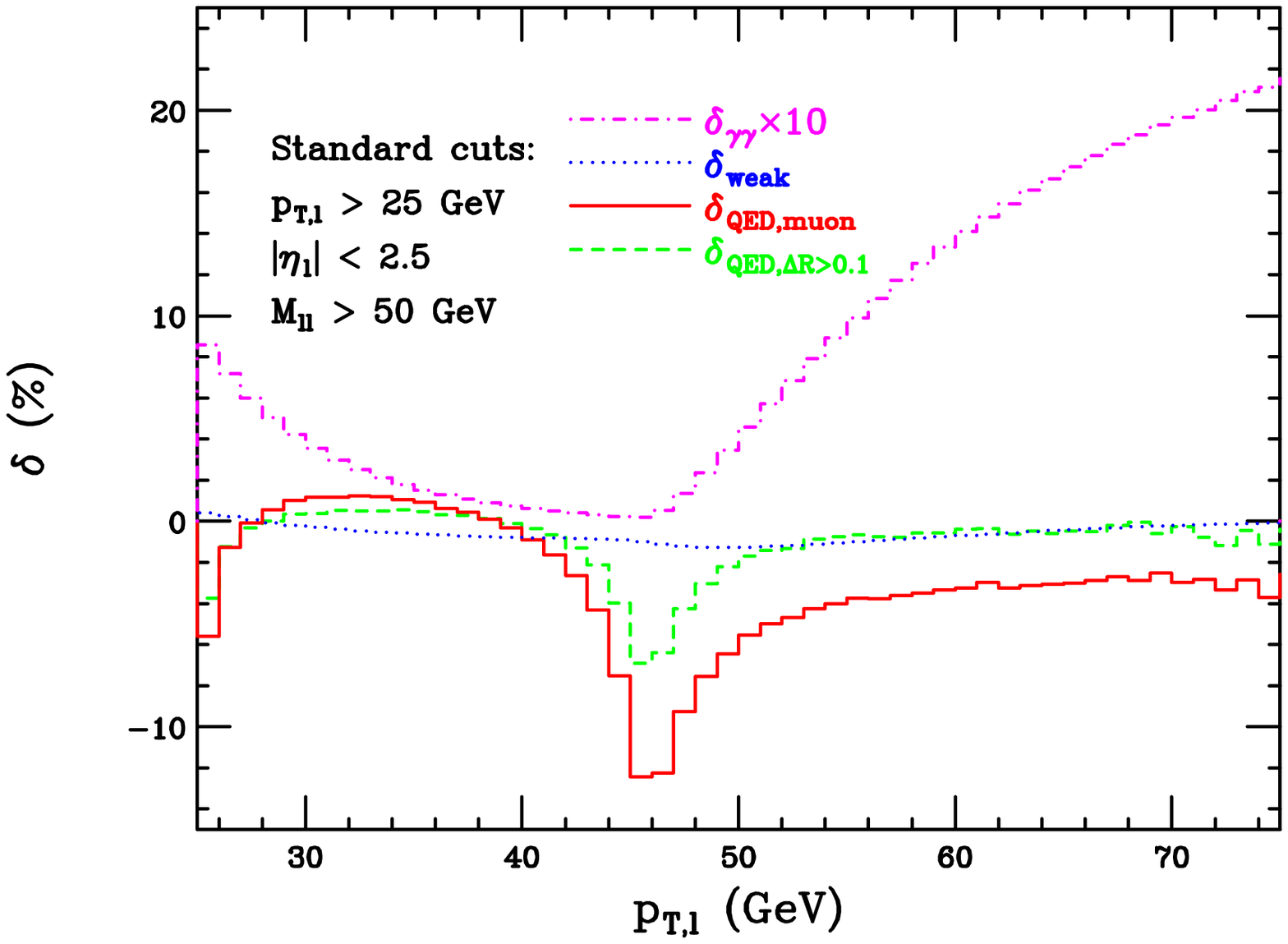}
\end{minipage}
\begin{minipage}[b]{3.0in}
\vspace{1cm}
  \includegraphics[width=3.0in]{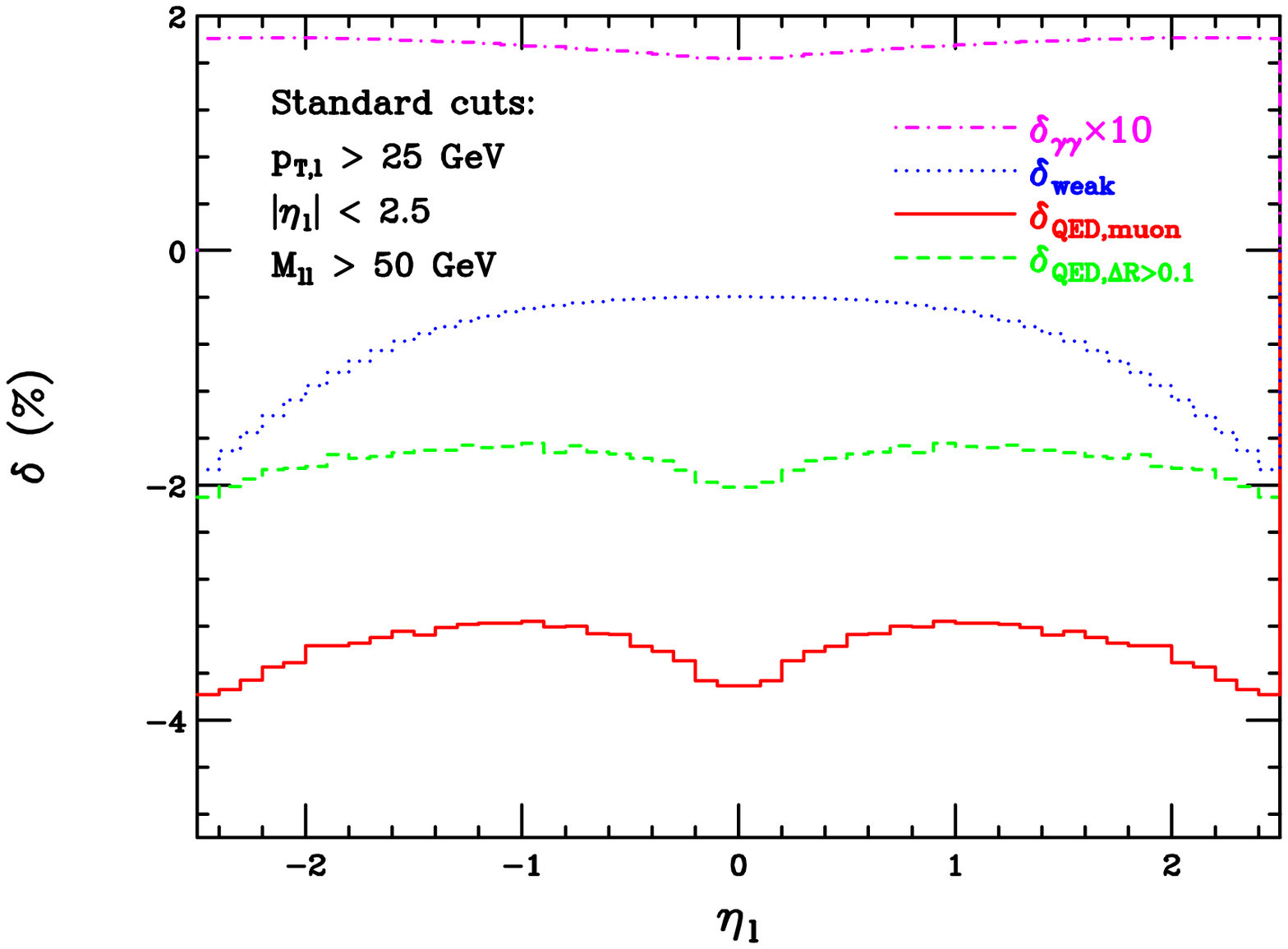}
\end{minipage}
\caption{Percentage change of various components of the NLO EW correction, with respect to the leading-order result.  The MRST2004QED set is used, and the LHC center-of-mass energy is taken to be 14 TeV.  Clockwise from the upper left, the plots show the lepton-pair invariant mass, the lepton-pair rapidity, the lepton $p_T$, and the lepton pseudorapidity}
\label{PercentageChange}
\end{figure}

Fig.~\ref{PercentageChange} uses the histogramming feature of FEWZ to reproduce the percent changes in various distributions arising from QED and weak corrections in the $Z$ resonance region. We show several different lepton identification scenarios. An ideal resolution is assumed in the bare muon case, and no recombination procedure is applied. For massless leptons, two values of separation $\Delta R_{l\gamma}$ are considered for the recombination procedure. A larger $\Delta R_{l\gamma}$ effectively lowers the detector resolution and leads to a more inclusive observable. The photonic correction is dominated by FSR, which is particularly sensitive to the choice of the recombination procedure.  

The invariant mass of the lepton pair $M_{ll}$ receives a positive photonic correction below the $Z$ peak and a negative correction above the $Z$ peak, leading to a shift of the $Z$ resonance position to a smaller value. The weak correction has a similar feature, but its size is an order of magnitude smaller in the resonance region. The lepton pT distribution has a well-known Jacobian peak at $M_Z/2$, which is also distorted primarily by the photonic correction.  We note that a proper description of the $p_{T,l}$ distribution near the Jacobian peak requires resummation of multiple soft photon effects.  The bottom two histograms in Fig.~\ref{PercentageChange} are the rapidity distributions of the reconstructed $Z$ and of the lepton. The NLO EW correction to the $Z$ rapidity is roughly constant except at the boundary of phase space. The contribution of photon-induced dilepton production is highly suppressed by the photon PDF and is negligible in most situations. It grows relatively more important in the high rapidity region due to the nature of its t-channel exchange diagrams. All histograms are consistent with the distribution shapes given in Ref.~\cite{Dittmaier:2009cr}.  The extensive comparison presented above reveals good agreement between our results and previous work.

\subsection{Combination of electroweak and QCD corrections in FEWZ}

\begin{figure}[!ht]
\begin{minipage}[b]{3.0in}
  \includegraphics[width=3.0in]{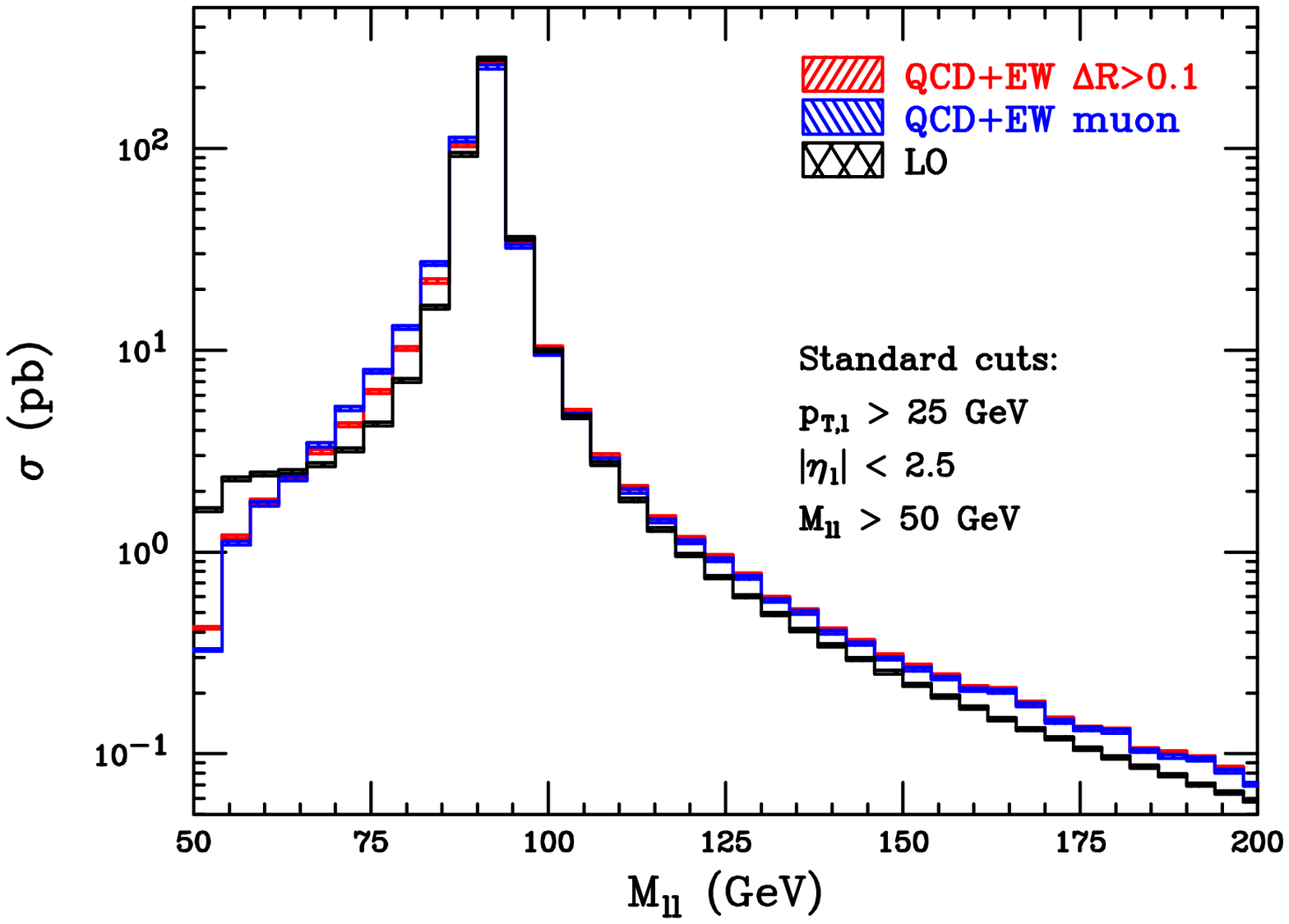}
\end{minipage}
\begin{minipage}[b]{3.0in}
  \includegraphics[width=3.0in]{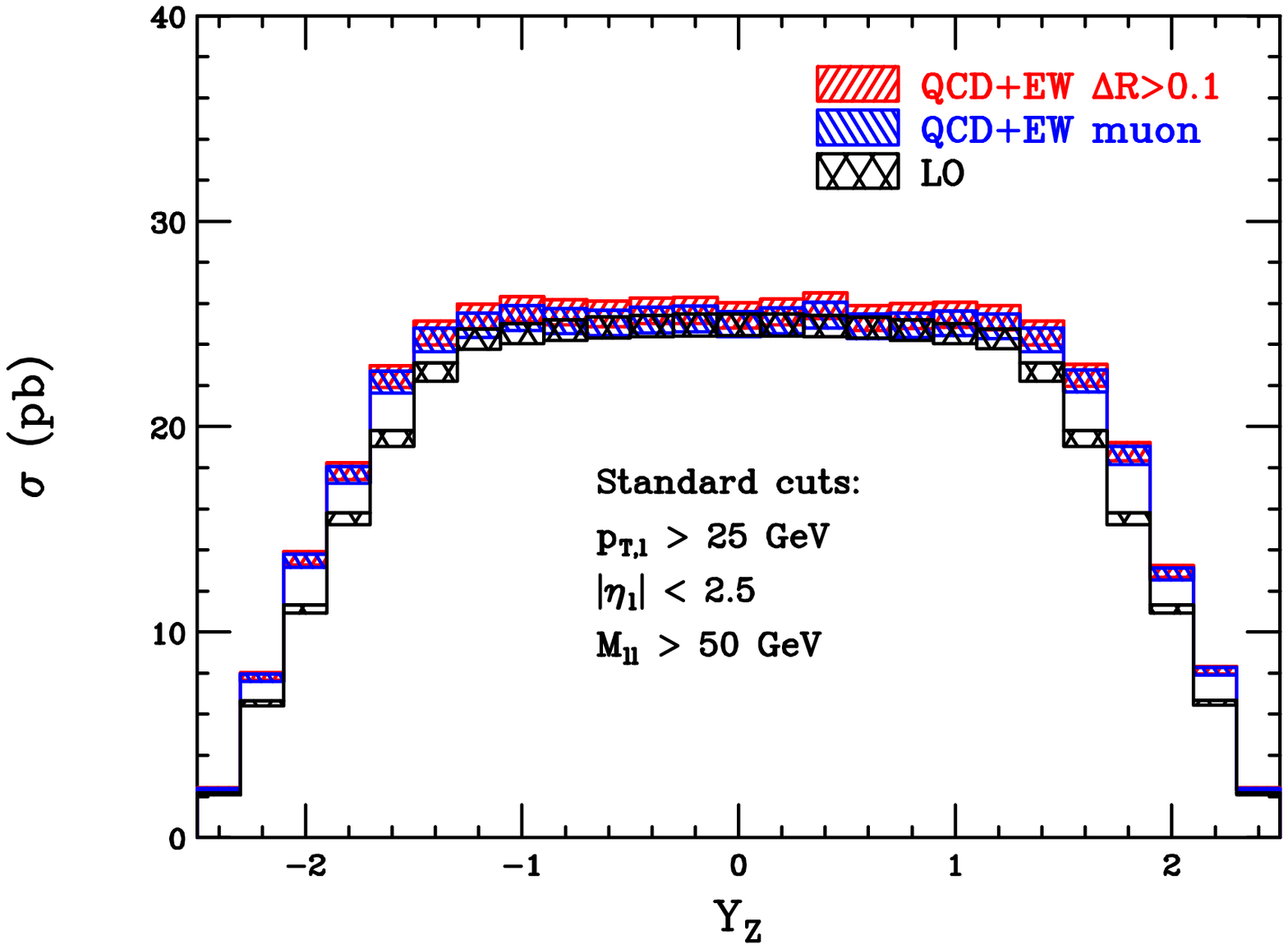}
\end{minipage}
\begin{minipage}[b]{3.0in}
\vspace{1cm}
  \includegraphics[width=3.0in]{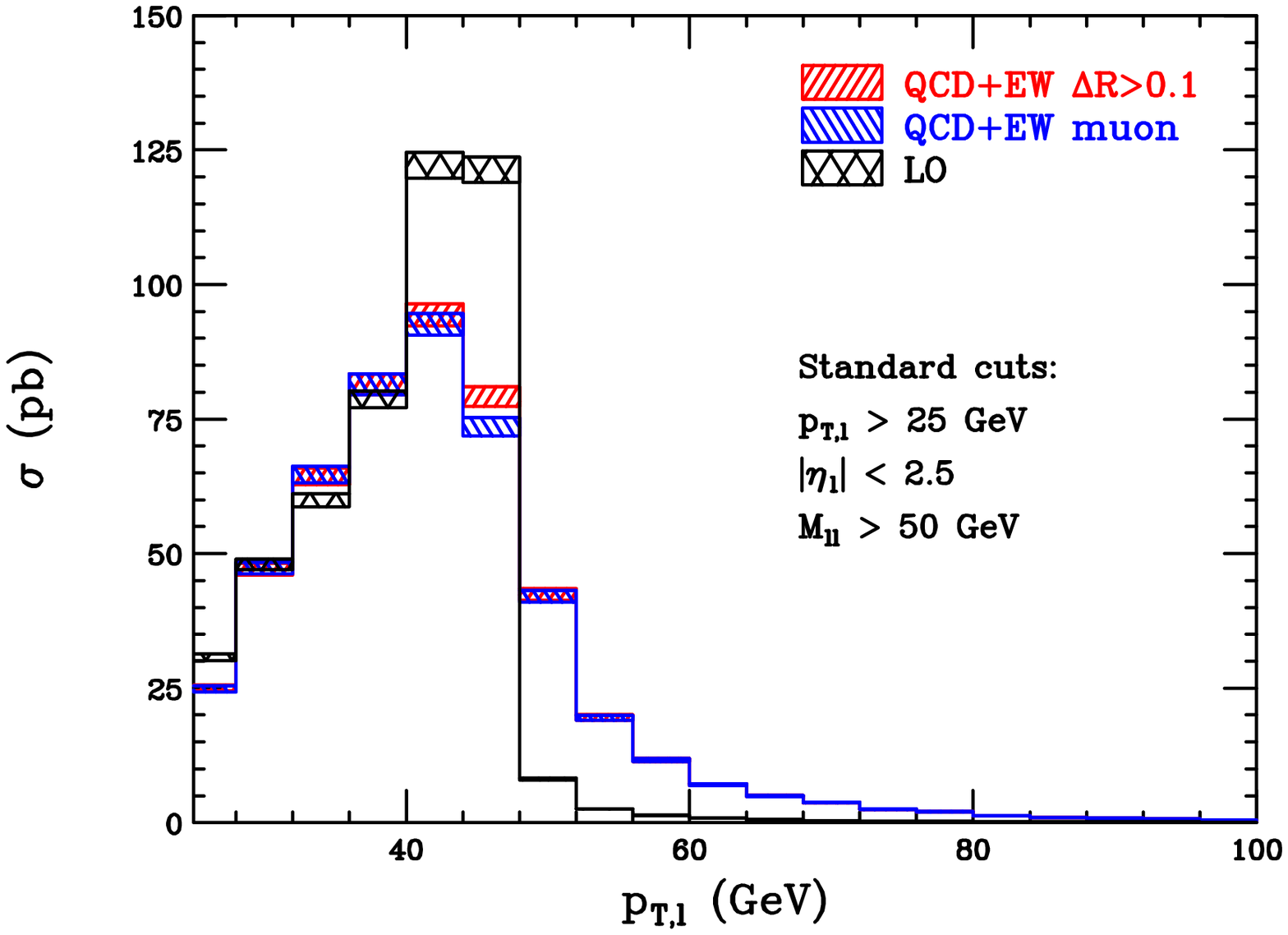}
\end{minipage}
\begin{minipage}[b]{3.0in}
\vspace{1cm}
  \includegraphics[width=3.0in]{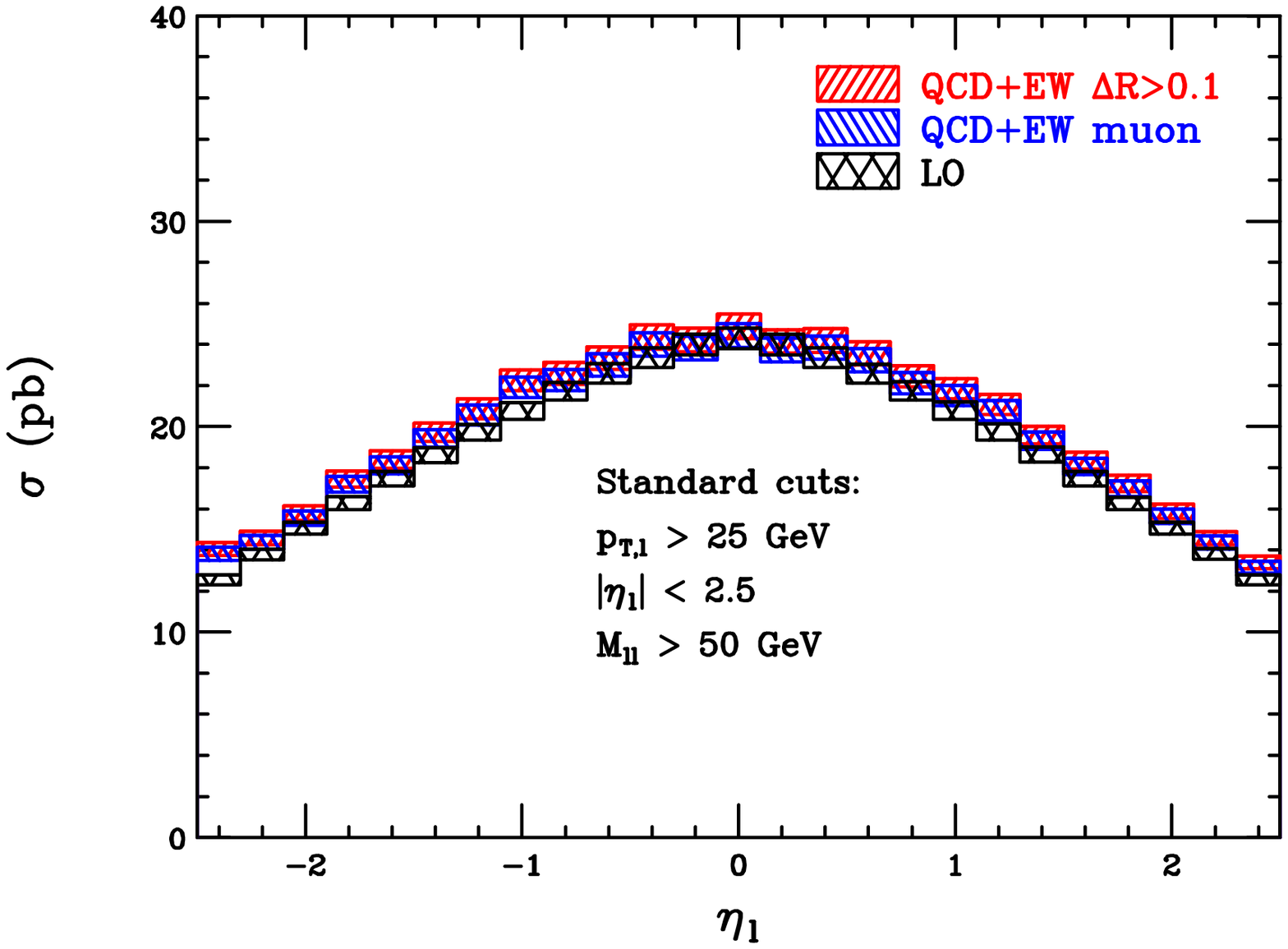}
\end{minipage}
\caption{Several representative distributions showing the combination of NNLO QCD and NLO EW corrections to lepton-pair production at a $\sqrt{s}=8$ TeV LHC.  The MSTW2008NNLO PDF set has been used, with the hatched regions corresponding to the estimated 68\% PDF error.  The $\alpha_S$ error has not been included.  Clockwise from the upper left, the plots show the lepton-pair invariant mass, the lepton-pair rapidity, the lepton $p_T$, and the lepton pseudorapidity.}
\label{AllNLOCorr}
\end{figure}

Having validated our calculation, we next study the combination of the NLO EW and the NNLO QCD corrections.  We switch to $\sqrt{s}=8$ TeV in this subsection.  In Fig.~(\ref{AllNLOCorr}), we display the full corrections to dilepton production for several observables as a demonstration of the full functionality of FEWZ. The NLO EW correction, as explained above, plays an important role in defining the distribution shapes near the Z resonance.  The shift of events from the $Z$-peak to lower invariant masses caused by FSR effects is apparent in the uppermost left plot.   The large EW Sudakov logarithms cause the rise in the high-mass tail of the invariant-mass distribution seen in the plot.  We note that because of the cut $p_{T,l}>25$ GeV, the invariant mass of the lepton pair is restricted to $M_{ll}> 50$ GeV, coinciding exactly with the cut on this variable that we impose.  Sensitivity to this phase-space restriction leads to the large shift from the leading-prediction near this boundary.  The lepton $p_T$ distribution in the lower-left panel exhibits the usual Jacobian 
peak at $M_Z/2$.  Resummation of soft-photon and soft-gluon effects is needed for a proper description near this boundary.
Both the lepton pseudorapidity and lepton-pair rapidity in the rich panels show little sensitivity to higher-order effects.

\subsection{Distributions of photon radiation in Drell-Yan production}

We proceed to examine distributions of photon radiation at the LHC.  To define photon experimentally, we impose the following cuts in our analysis:
\begin{equation}
 p_{T,\gamma} > 20\, \text{GeV}, \;\;\; |\eta_{\gamma}| < 2.5, \;\;\; \Delta R_{l\gamma}>0.05.
 \label{cuts1}
\end{equation}
We note that the experimental capabilities permit a much lower cut on photon $p_T$, reaching down to 5 GeV, but we set 
the cut high here to reveal certain kinematic features in the prediction.  We focus on the invariant-mass region below the $Z$-peak, where FSR effects are enhanced.  The following cuts are also 
applied in our study:
\begin{eqnarray}
p_{T,l}^{hard} &>& 30 \,\text{GeV}, \;\;\; 30 \,\text{GeV} < M_{ll} < 86 \, \text{GeV}, \nonumber \\
p_{T,l}^{soft} &>& 10 \,\text{GeV}, \;\;\; |\eta_l| < 2.5.
 \label{cuts2}
\end{eqnarray}
These cuts are motivated by an ongoing study of photon radiation in the Drell-Yan process within CMS~\cite{cms}.  We also study distributions where the softer-lepton $p_T$ cut is increased to $p_{T,l}^{soft} > 20 \,\text{GeV}$.  To facilitate comparison with the ongoing experimental study we consider the $\sqrt{s}=7$ TeV LHC.

\begin{figure}[!ht]
\begin{minipage}[b]{3.0in}
  \includegraphics[width=3.0in]{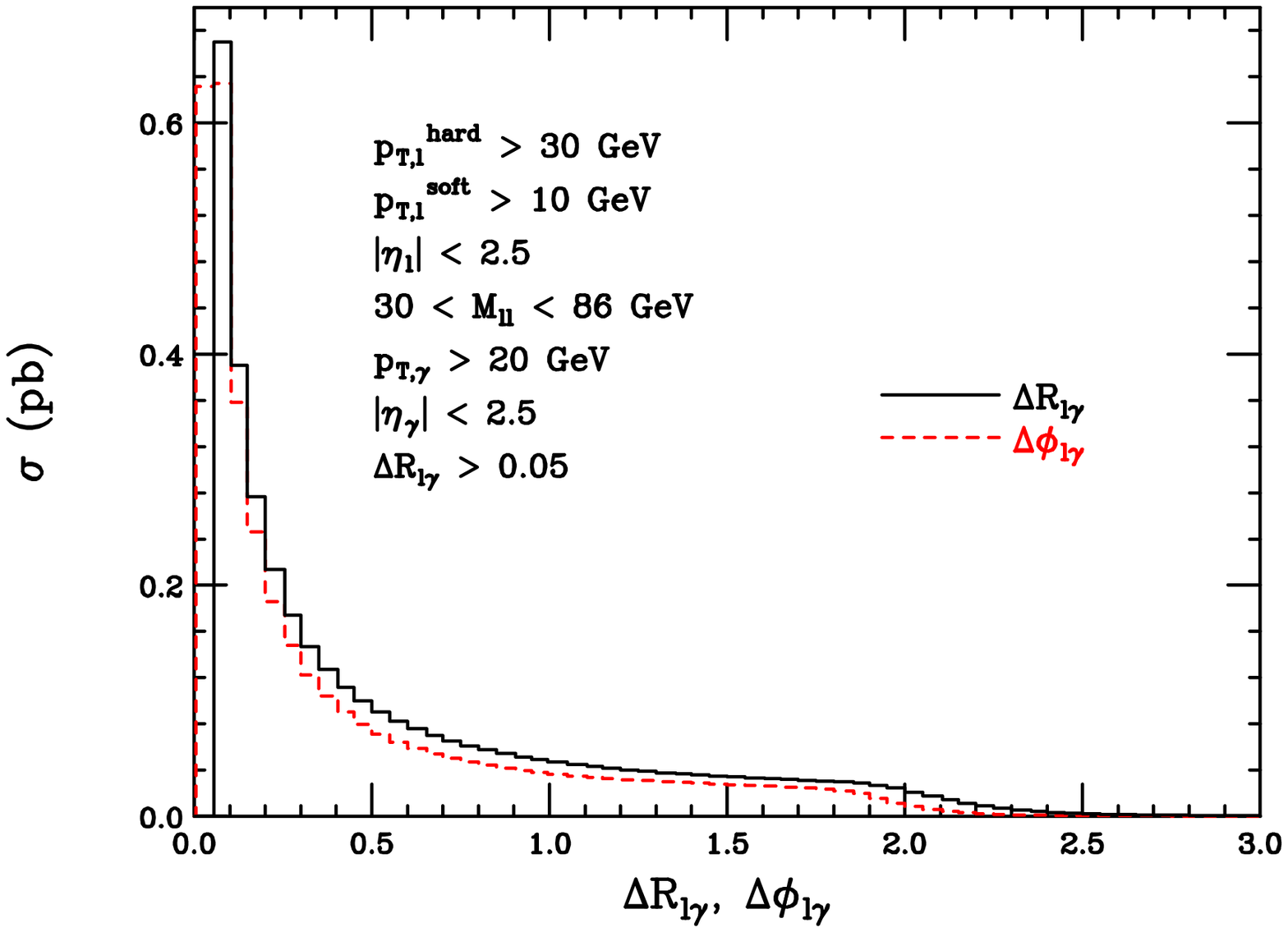}
\end{minipage}
\begin{minipage}[b]{3.0in}
    \includegraphics[width=3.0in]{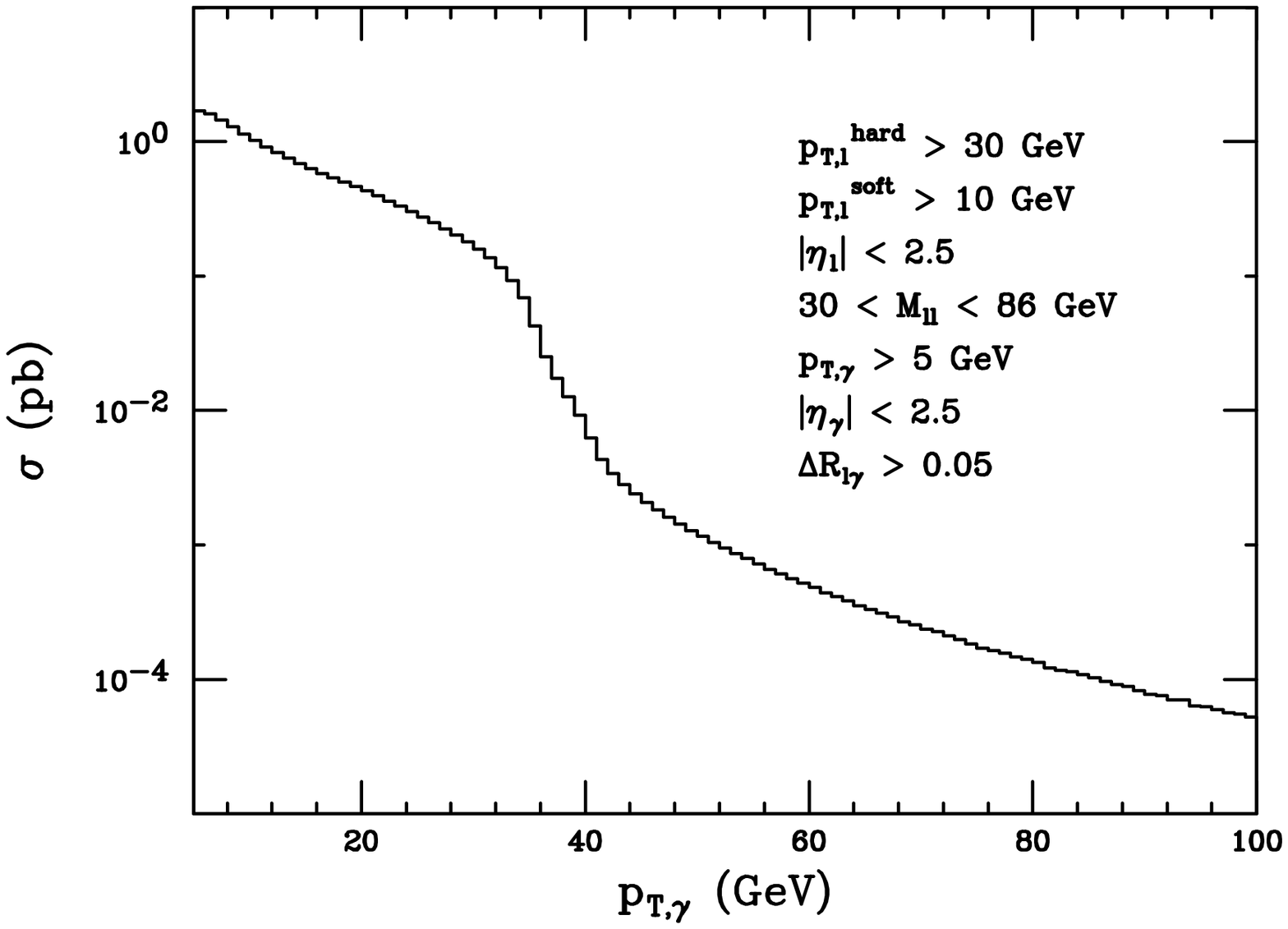}
\end{minipage}
\caption{Distributions of photon radiation in the Drell-Yan process.  The left panel shows the distribution of separation between the photon and the nearest leptons using both $\Delta \phi$ and $\Delta R$ as a distance measure.  The right panel shows the photon transverse momentum spectrum.  The cuts are those described in Eqs.~(\ref{cuts1}) and~(\ref{cuts2}).}
\label{PhotonDist}
\end{figure}

\begin{figure}[ht]
  \includegraphics[width=4.0in]{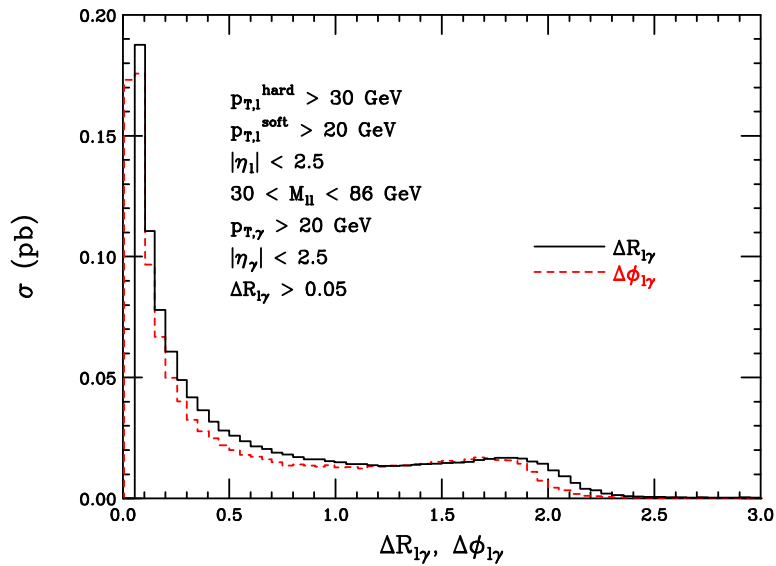}
\caption{The separation between the photon and the nearest lepton using instead $p_{T,l}^{soft} > 20 \,\text{GeV}$.}
\label{PhotonDist2}
\end{figure}

The left histogram in Fig.~\ref{PhotonDist} shows the distribution of $\Delta R_{l\gamma} =  \sqrt{(\Delta \eta_{l\gamma})^2+ (\Delta \phi_{l\gamma})^2}$, defined as the separation between the photon and the nearest lepton. It peaks toward zero separation, corresponding to the FSR collinear divergence. We also present the distribution of $\Delta \phi_{l\gamma}$, the difference in azimuthal angle between the photon and the nearest lepton, in the same histogram. There is a kink in both the $\Delta R_{l\gamma}$ and $\Delta \phi_{l\gamma}$ distributions, the location of which is roughly given by $\pi-\arccos(p_{T,\gamma,min}/M_Z) \approx \pi/2$ for small $p_{T,\gamma,min}$.  It occurs for the maximum $\Delta \phi_{l\gamma}$ when the two leptons and the photon travel with the same pseudorapidity.  The exact location of the feature is at the angle $\pi-\arccos(p_{T,\gamma,min}/\sqrt{M_Z^2+p^{2}_{T,\gamma,min}})$ for ISR photons and $\pi-\arccos(p_{T,\gamma,min}/(M_Z-p_{T,\gamma,min}))$ for FSR photons. At small $p_{T,\gamma,min}$, the two expressions yields roughly the same number.  However, the dominance of FSR photons with small separation washes out this effect.  At large $p_{T,\gamma,min}$, the collinear FSR contribution becomes suppressed and the prominence of the effect increases.  We can enhance the kinematic feature by increasing the cut on the softer lepton, which further reduces the collinear FSR 
contribution.  This is shown in Fig.~\ref{PhotonDist2}, where we have used $p_{T,l}^{soft} > 20 \,\text{GeV}$.  The kink has now become a small peak.

The photon $p_T$ distribution in the right panel of Fig.~\ref{PhotonDist} has a shoulder at around 35 GeV. It can be explained by noting that the majority of photon radiation occurs as a result of bremsstrahlung radiation off a lepton from an on-shell $Z$. Most of the photon radiation is collinear to the direction of the lepton, and is limited to a maximum $p_T$ of half of the $Z$ mass as it recoils against the other lepton. However, the requirement of a lepton with a $p_T$ greater than 10 GeV reduces the maximally allowed $p_T$ for FSR photon to $M_Z/2-p_{T,l,min} \approx35\textrm{GeV} $ where $p_{T,l,min} = 10~\textrm{GeV}$.  We can interpret the small bump as the ``Jacobian peak" of the photon, in analogy to the lepton case. 

\section{Conclusion \label{sec:conc}}

In this paper we have presented a combination of the NNLO QCD and NLO EW corrections to the Drell-Yan production of a lepton pair.  This combination was not previously available in the literature.  We have incorporated these 
corrections into an updated version of the analysis code FEWZ, which is used in numerous experimental studies.  The 
combination of both the QCD and EW sources of higher-order corrections in a single program eliminates the need to piece 
them consistently.  We have recalculated the EW corrections in the complex-mass scheme and have cross-checked our results in detail against previous results in the literature.  We have presented several new techniques in the course of 
our calculation, including a combination of phase-space slicing and subtraction methods for handling real-emission 
corrections, and a treatment for the analytic continuation of two-point functions needed in the complex-mass scheme.

We have presented numerous phenomenological results relevant for LHC studies, including the first theoretical results 
for distributions containing both NNLO QCD and NLO EW effects.  We have also discussed several interesting kinematic features in photon-radiation distributions caused by imposing experimental cuts.  These should be observable in the LHC data, and would offer interesting tests of the underlying mechanism for producing photons in association with a lepton pair in hadronic collisions.

\section*{Acknowledgements}

\noindent
We are grateful to A.~Kubik, M.~Schmitt and S.~Stoynev for useful discussions on the ongoing experimental studies at 
CMS.  We also thank S. Quackenbush for advice and help in finalizing the new version of FEWZ.  This work is supported by the U.S. Department of Energy, Division of High Energy Physics, under contract DE-AC02-06CH11357 and the grants DE-FG02-95ER40896 and DE-FG02-08ER4153, and with funds provided by Northwestern University.

\appendix

\section{Analytic Continuation of the Scalar Bubble Integral \label{sect:acB0}}

The two-point one-loop scalar integral (the bubble integral) is defined as follows:

\begin{equation}
B_0(s;M_1^2,M_2^2)  =
 \frac{(2\pi\mu)^{4-D}}{i \pi^2}\int d^D l \;
 \frac{1}
{(l^2-M_1^2+i\varepsilon)
((l+p)^2-M_2^2+i\varepsilon)}\,,
 \end{equation}

\noindent
where $s=p^2$. The space-time dimension is taken to be $D=4-2\epsilon$.  Introducing a standard Feynman parameterization of the integral, we obtain

\begin{eqnarray}
B_0(s;M_1^2,M_2^2)  
&=& (4\pi\mu^2)^{\epsilon} \Gamma(\epsilon) \int_0^1 \; dx \; 
[-x (1-x) s +x M_2^2+(1-x) M_1^2 -i \varepsilon]^{-\epsilon}  \nonumber \\
&=& (4\pi\mu^2)^{\epsilon}  
\left\{ \frac{1}{\epsilon}
 -\int_0^1 \; dx \; \ln (-x(1-x) s +x M_2^2+(1-x) M_1^2 -i \varepsilon ) \right\}
 +\mathcal{O}(\epsilon) . \nonumber\\
 \end{eqnarray}

\begin{figure}[h]
\begin{center}
\includegraphics[scale=0.6,angle=0]{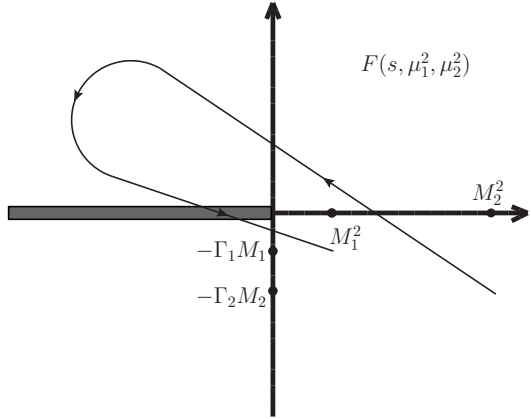}
\caption{The contour for $\text{ln}\,F$ in the complex plane, assuming arbitrary complex parameters.}
\label{complexbubble}
\end{center}
\end{figure}

The logarithm is defined to have a branch cut on the negative real axis. If $s$, $M_1^2$, and $M_2^2$ are real parameters, the integration contour stays below the branch cut due to the Feynman prescription for the propagators. If we replace $M_1^2$ and $M_2^2$ with the complex masses $\mu_{1,2}^2=M_{1,2}^2-i\Gamma_{1,2}M_{1,2}$, the contour simply moves further away from the branch cut. This can change if $s$ is also a complex mass, as shown in Fig.~\ref{complexbubble}.  If the masses $M_1$ and $M_2$ are sufficiently large so that both particles in the loop cannot go on shell, then the argument inside the logarithm of the integrand always has a positive real part, and the contour never crosses the branch cut.  If the particles in the loop can go on-shell, the branch cut can be crossed.

For the computation of the gauge boson self-energy insertions, the particles running in the loop are either massless fermions, a heavy top quark, a Higgs boson or the gauge boson themselves. Except for massless fermions, the other internal particles can never be on-shell. Therefore, we only need to worry about the bubble integral of the form $B_0(s,0,0)$ with $s=\mu_Z^2$ or $\mu_W^2$. It is easy to work out the integral assuming s is the complex mass of either the $W$ or $Z$ boson:
\begin{eqnarray}
B_0(s;0,0)  
&=& (4\pi\mu^2)^{\epsilon}  
\left\{ \frac{1}{\epsilon}
 -\int_0^1 \; dx \; \ln (-x(1-x) s -i \varepsilon ) \right\}
 +\mathcal{O}(\epsilon)\nonumber\\
 &=& (4\pi\mu^2)^{\epsilon}  
\left\{ \frac{1}{\epsilon} + 2 
 - \left[ \ln (s-i\varepsilon) +\ln(-1-i\varepsilon/s)\right] \right\}
 +\mathcal{O}(\epsilon) .
 \end{eqnarray}
In the last line, we have written the integral in a form valid for both real $s$, or if $s$ is given by a complex mass.  The second logarithm consistently yields $-i\pi$ as long as $\textrm{Re}(s) > 0$, ensuring a smooth transition for a complex mass valued $s$.  It is easy to find the following Taylor expansion assuming a small imaginary part for $s$:
 \begin{equation}
B_0(\mu_{W/Z}^2;0,0) =  B_0(M_{W/Z}^2;0,0) + (\mu_{W/Z}^2-M_{W/Z}^2) \left.\frac{d B_0(s;0,0)}{ds}\right|_{s=M_{W/Z}^2}  + \mathcal{O}(\Gamma_{W/Z}^2/M_{W/Z}^2) ,
 \end{equation}
which demonstrates the equivalence at NLO between our result and the alternative procedure proposed in Ref.~\cite{Denner:2005fg}.

\end{document}